\documentstyle[aps,preprint]{revtex}
\begin{document}
\title{A supersymmetric mean-field approach to the infinite-$U$ Hubbard model}
\author{V.Yu.Irkhin and A.A.Katanin}
\address{Institute of Metal Physics, 620219 Ekaterinburg, Russia.}
\maketitle

\begin{abstract}
A generalized supersymmetric representation of the Hubbard operator algebra
is considered. This representation is applied to the infinite-$U$ Hubbard
model. A mean-field theory which takes into account both on-site and
inter-site virtual boson-fermion transitions is developed. Unlike previous
approaches, the mean-field theory considered is free from divergences. A
possible application of these results to the ferromagnet-paramagnet
transition, as well as to other problems is discussed.
\end{abstract}

\section{Introduction}

The description of magnetism within the Hubbard model is the long-standing
problem. At $U=\infty $ the ferromagnetic (but not antiferromagnetic) state
can occur, and, as shown by Nagaoka, for small hole concentrations $\delta $
this is the ground state. The stability of saturated ferromagnetism for not
too small $\delta $ was investigated within different approaches (see, e.g.
Refs. \cite{Roth,Edw,Nolting,ZI}). The critical concentration of holes was
estimated as $\delta _c\simeq 1/3$ for the transition into non-saturated
ferromagnetic state, and as $\delta _c^{\prime }\simeq 2/3$ for that into
non-magnetic one. The same result for $\delta _c$ can be obtained by
comparing the total energies in the slave-fermion approach (which describes
correctly the excitations near half-filling) and slave-boson approach (which
is suitable for the non-magnetic phase), see Ref. \cite{sb-sf} and
references therein.

Slave-fermion and slave-boson approaches have the advantage that they
correspond to the $N\rightarrow \infty $ limit of the generalized $U=\infty $
Hubbard model with $N$ fermion flavors (which will be referred to as $%
SU(N|1) $ model), and $1/N$ corrections can be found in a regular way. These
two approaches treat different kind of excitations, and therefore have
different applicability regions. While slave-fermion approach is convenient
at small hole concentrations and enables one to describe naturally the
magnetic state, the slave-boson approach works well at high $\delta $.

To obtain the physical picture at intermediate hole concentrations, a
supersymmetric approach should be developed which interpolates between
slave-fermion and slave-boson ones. The situation here is similar to that in 
$s-f$ model where boson representation of impurity spin operators describes
correctly magnetic phases, while fermion representation is suitable for
description of Kondo (i.e. nonmagnetic) state. To describe continuous
transition between these phases, the supersymmetric approach of Ref.\cite
{Coleman} can be used.

In this paper we consider the application of the supersymmetric approach to
the $SU(N|1)$ model and obtain the effective action of the infinite-$U$
Hubbard model in supersymmetric representation. Then we consider the
mean-field approach to this action and obtain the corresponding
self-consistent equations.

\section{Representations of the Hubbard algebra}

The standard way of treating large-$U$ Hubbard model is introducing the
Hubbard operators $X^{\alpha \beta }$ ($\alpha ,\beta =0,\pm $). These
satisfy at a lattice site the commutation relations 
\begin{equation}
\lbrack X^{\alpha \beta },X^{\gamma \delta }]_{\pm }=\delta _{\beta \gamma
}X^{\alpha \delta }\pm \delta _{\alpha \delta }X^{\gamma \beta }
\label{Comm}
\end{equation}
and the constraint 
\begin{equation}
\sum_\alpha X^{\alpha \alpha }=1  \label{C1}
\end{equation}
The operators $X^{\alpha \beta }$ give the spin $S=1/2$ realization of the $%
SU(2|1)$ superalgebra. To construct $1/N$ expansion we have to generalize
this algebra to $SU(N|1)$ by introducing the operators $\chi ^{\alpha \beta
} $ ($\alpha ,\beta =0...N$) which satisfy the same commutation relations (%
\ref{Comm}). We also use the generalized form of the constraint 
\begin{equation}
\sum_\alpha \chi ^{\alpha \alpha }=Q_0
\end{equation}
which reduces to standard one, (\ref{C1}), for the physical case $Q_0=1$.
The operators $\chi ^{\alpha \alpha }$ yield a representation of the $%
SU(N|1) $ superalgebra with the ``superspin'' $Q_0/N$.

The slave-fermion representation of the operators $\chi ^{\alpha \beta }$
through $N$ bosons and one fermion has the form 
\begin{equation}
\chi ^{\sigma \sigma ^{\prime }}=b_\sigma ^{\dagger }b_{\sigma ^{\prime
}},\;\chi ^{\sigma 0}=b_\sigma ^{\dagger }f,\;\chi ^{00}=f^{\dagger }f
\label{sf}
\end{equation}
($\sigma ,\sigma ^{\prime }=1...N$), with $b_\sigma $ and $f$ being Bose and
Fermi operators respectively. For $Q_0<N$ there exists also the slave-boson
representation 
\begin{equation}
\chi ^{\sigma \sigma ^{\prime }}=c_\sigma ^{\dagger }c_{\sigma ^{\prime
}},\;\chi ^{\sigma 0}=c_\sigma ^{\dagger }a,\;\chi ^{00}=a^{\dagger }a
\label{sb}
\end{equation}
where $c_\sigma $ and $a$ are Fermi and Bose operators.

To obtain the supersymmetric representation which interpolates between
slave-fermion and slave-boson ones, we introduce, following to Refs. \cite
{Coleman,Coleman1,IK}, the operators 
\begin{equation}
\Psi _\sigma =\left( 
\begin{array}{c}
c_\sigma \\ 
b_\sigma
\end{array}
\right) ,\;\Psi _0=\left( 
\begin{array}{c}
a \\ 
f
\end{array}
\right)
\end{equation}
($\sigma =1...N$) with 
\begin{eqnarray}
\chi ^{\alpha \beta } &=&\Psi _\alpha ^{\dagger }\Psi _\beta ,  \nonumber \\
Q_0 &=&\Psi _\alpha ^{\dagger }\Psi _\alpha
\end{eqnarray}
To make a distinction between the representations with different symmetry,
we consider the second-order Casimir operator of $SU(N|1)$%
\begin{equation}
{\cal C}_2=\chi ^{\sigma \sigma ^{\prime }}\chi ^{\sigma ^{\prime }\sigma
}-\chi ^{\sigma 0}\chi ^{0\sigma }+\chi ^{0\sigma }\chi ^{\sigma 0}-(\chi
^{00})^2  \label{C2}
\end{equation}
Expressing this in terms of $\Psi $ we obtain 
\begin{equation}
{\cal C}_2=\Psi _\alpha ^{\dagger }\Psi _\alpha [N-1-\Psi _\beta ^{\dagger
}\tau _3\Psi _\beta ]-[\Theta ,\Theta ^{\dagger }]
\end{equation}
where $\tau _3$ is the Pauli matrix, 
\begin{equation}
\Theta =b_\sigma ^{\dagger }c_\sigma -f^{\dagger }a
\end{equation}
is the mixing fermion-boson operator with $\{\Theta ,\Theta ^{\dagger
}\}=Q_0 $, and the summation over repeated indices is assumed. Taking into
account the constraint (\ref{C2})\ we have finally 
\begin{equation}
{\cal C}_2=Q_0(N-1-Y)
\end{equation}
where 
\begin{equation}
Y=\Psi _\beta ^{\dagger }\tau _3\Psi _\beta +\frac 1{Q_0}[\Theta ,\Theta
^{\dagger }]
\end{equation}
By fixing $Y=-Q_0+1$ ... $Q_0-1$, we obtain representations with different
symmetry.

\section{$SU(N|1)$ generalization of the $U=\infty $ Hubbard model}

The Hamiltonian of the $SU(N|1)$ model can be now rewritten as 
\begin{equation}
{\cal H}=\sum_{ij}t_{ij}\chi _i^{\sigma 0}\chi _j^{0\sigma
}=\sum_{ij}t_{ij}\Psi _{i\sigma }^{\dagger }\Psi _{i0}\Psi _{j0}^{\dagger
}\Psi _{j\sigma }  \label{HtJ}
\end{equation}
For $N=2,$ $Q_0=1$ this coincides with that of the $U=\infty $ Hubbard
model. The partition function reads 
\begin{equation}
{\cal Z}=\int D\Psi \exp \left\{ -\int\limits_0^\beta d\tau ({\cal L}_0+%
{\cal H})\right\}  \label{Z}
\end{equation}
where 
\begin{equation}
{\cal L}_0=\sum_i\Psi _{i\alpha }^{\dagger }\left( \frac \partial {\partial
\tau }+\lambda _i+\mu \delta _{\alpha 0}+\zeta _i\tau _3\right) \Psi
_{i\alpha }-\frac{2\zeta }{Q_0}\sum_i\Theta _i^{\dagger }\Theta
_i-\sum_i(\zeta _iY+\lambda _iQ_0+\mu \delta )
\end{equation}
is the free Lagrangian, $\delta $ is the concentration of holes. Following
to Ref. \cite{Coleman}, we perform the replacement $\Psi \rightarrow g\Psi $
with 
\begin{equation}
g=\left( 
\begin{array}{cc}
1+\overline{\eta }\eta /2 & \eta \\ 
-\overline{\eta } & 1-\overline{\eta }\eta /2
\end{array}
\right)
\end{equation}
to obtain the gauge-invariant Lagrangian in the form 
\begin{eqnarray}
{\cal L}_0^{\prime } &=&\sum_i\Psi _{i\alpha }^{\dagger }(\partial _\tau
+\lambda _i+\mu \delta _{\alpha 0}+\zeta _i\tau _3)\Psi _{i\alpha } 
\nonumber \\
&&\ \ \ \ \ -\frac 1{Q_0}\sum_i\Theta _i^{\dagger }(\partial _\tau +2\zeta
_i)\Theta _i-\sum_i(\zeta _iY+\lambda _iQ_0+\mu \delta )
\end{eqnarray}
Performing the Hubbard-Stratonovich transformation, we obtain \cite{Coleman} 
\begin{eqnarray}
{\cal L}_0^{\prime } &=&\sum_i\Psi _{i\alpha }^{\dagger }\left[ \partial
_\tau +\lambda _i+\mu \delta _{\alpha 0}+\left( 
\begin{array}{cc}
\zeta _i & D_0\alpha _i \\ 
\overline{D}_0\alpha _i^{\dagger } & -\zeta _i
\end{array}
\right) \right] \Psi _{i\alpha }  \label{L0} \\
&&\ \ \ \ +Q_0\sum_i\alpha _i^{\dagger }D_0\alpha _i-\sum_i(\zeta
_iY+\lambda _iQ_0+\mu \delta )  \nonumber
\end{eqnarray}
with $D_0=\partial _\tau +2\zeta _i.$ The Hamiltonian (\ref{HtJ}) is
decoupled in the same way as in Ref.\cite{AL}, and we derive 
\begin{eqnarray}
{\cal H} &=&\sum_{ij}t_{ij}\left[ {\cal A}_{ij}c_{i\sigma }^{\dagger
}c_{j\sigma }+{\cal C}_{ij}a_ia_j^{\dagger }+{\cal F}_{ij}b_{i\sigma
}^{\dagger }b_{j\sigma }+{\cal B}_{ij}f_j^{\dagger }f_i\right.  \nonumber \\
&&\ \ \ \ \ \ +\overline{{\cal P}}_{ij}a_if_j^{\dagger }+f_ia_j^{\dagger }%
{\cal P}_{ji}+c_{i\sigma }^{\dagger }b_{j\sigma }{\cal Q}_{ji}+\overline{%
{\cal Q}}_{ij}b_{i\sigma }^{\dagger }c_{j\sigma }  \nonumber \\
&&\ \ \ \ \ \ \left. -{\cal A}_{ij}{\cal C}_{ji}+{\cal B}_{ij}{\cal F}_{ji}-%
\overline{{\cal P}}_{ij}{\cal Q}_{ji}-\overline{{\cal Q}}_{ij}{\cal P}%
_{ji}\right]  \nonumber \\
\ &=&\sum_{ij}t_{ij}\left[ \Psi _{i\sigma }^{\dagger }{\cal V}_{ij}\Psi
_{j\sigma }+\Psi _{i0}^{\dagger }{\cal Z}_{ij}\Psi _{j0}-\text{STr}({\cal V}%
_{ij}{\cal Z}_{ji})\right]  \label{H}
\end{eqnarray}
where STr$(...)$ is the supertrace; ${\cal P},\overline{{\cal P}},{\cal Q},%
\overline{{\cal Q}}$ are independent Grassmann variables, 
\begin{equation}
{\cal C}_{ij}=-{\cal A}_{ij}^{\dagger },\;{\cal B}_{ij}={\cal F}%
_{ij}^{\dagger }
\end{equation}
and 
\begin{equation}
{\cal V}_{ij}=\left( 
\begin{array}{cc}
{\cal A}_{ij} & {\cal Q}_{ji} \\ 
\overline{{\cal Q}}_{ij} & {\cal F}_{ij}
\end{array}
\right) ,\;{\cal Z}_{ij}=\left( 
\begin{array}{cc}
{\cal C}_{ij} & -{\cal P}_{ji} \\ 
-\overline{{\cal P}}_{ij} & {\cal B}_{ij}
\end{array}
\right)
\end{equation}
The model (\ref{Z}) with (\ref{L0}) and (\ref{H}) is invariant under the
gauge transformation 
\begin{eqnarray}
\Psi _i &\rightarrow &g_i\Psi _i  \nonumber \\
\alpha &\rightarrow &\alpha +(\partial _\tau +2\zeta _i)\eta _i  \nonumber \\
{\cal V}_{ij} &\rightarrow &g_i{\cal V}_{ij}g_j^{-1},\;{\cal Z}%
_{ij}\rightarrow g_i{\cal Z}_{ij}g_j^{-1}  \label{G}
\end{eqnarray}
The action ${\cal S}={\cal L}_0^{\prime }+{\cal H}$ can be rewritten in the
form 
\begin{eqnarray}
{\cal S} &=&\sum_i\Psi _{i\sigma }^{\dagger }\left[ (\partial _\tau +\lambda
_i+\zeta _i\tau _3)\delta _{ij}+\left( 
\begin{array}{cc}
{\cal A}_{ij}t & Q_{ji} \\ 
\overline{Q}_{ij} & {\cal F}_{ij}t
\end{array}
\right) \right] \Psi _{j\sigma }  \label{Sf} \\
&&\ \,\,+\sum_i\Psi _{i0}^{\dagger }\left[ (\partial _\tau +\lambda _i+\mu
+\zeta _i\tau _3)\delta _{ij}+\left( 
\begin{array}{cc}
{\cal C}_{ij}t & P_{ji} \\ 
\overline{P}_{ij} & {\cal B}_{ij}t
\end{array}
\right) \right] \Psi _{j0}  \nonumber \\
&&\ \ \ \ \ +Q_0\sum_i\alpha _i^{\dagger }D_0\alpha _i-\text{STr}({\cal V}%
_{ij}{\cal Z}_{ji})-\sum_i(\zeta _iY+\lambda _iQ_0+\mu \delta )  \nonumber
\end{eqnarray}
where 
\begin{eqnarray}
Q_{ji} &=&D_0\alpha _i\delta _{ij}+t{\cal Q}_{ji},\;\overline{Q}%
_{ij}=D_0\alpha _i\delta _{ij}+t{\cal Q}_{ji}  \nonumber \\
P_{ji} &=&D_0\alpha _i\delta _{ij}-t{\cal P}_{ji},\;\overline{P}_{ij}=%
\overline{D}_0\alpha _i^{\dagger }\delta _{ij}-t\overline{{\cal P}}_{ij}
\end{eqnarray}

\section{Mean-field theory}

Our purpose now is to consider the mean-field approximation for the action (%
\ref{Sf}), which does not violate the gauge invariance (\ref{G}). To this
end, we will take into account the fluctuations of ${\cal Q},{\cal P}$ and $%
\alpha ,$ while other fields will be considered in the mean-field
approximation. The motivation for this approximation is as follows.

(i) The fields ${\cal Q},{\cal P}$ and $\alpha $ are equally important:
while the field $\alpha $ describes the on-site virtual transitions of
bosons into fermions and vice versa, the fields ${\cal Q}$ and ${\cal P}$
describe the same transitions with simultaneous intersite hopping. As can be
seen from (\ref{Sf}), ${\cal Q}$ and ${\cal P}$ play the role of ``spatial
components'' of the gauge field, while $\alpha $ is only its time component.

(ii) All the fields, besides ${\cal Q},{\cal P}$ and $\alpha ,$ being taken
into account at the mean-field level, are shifted properly by the gauge
transformation. At the same time, the fields ${\cal Q},{\cal P}$ and $\alpha 
$ have zero mean-field value due to their fermionic nature and therefore can
not be transformed at the mean-field level.

Thus, ${\cal Q},{\cal P}$ and $\alpha $ make the minimal set of fields,
fluctuations of which should be taken into account to keep gauge invariance.
The gauge transformation of resulting theory is considered in Appendix. Note
that taking into account only fluctuations of the field $\alpha $ leads to
divergences due to violation of gauge invariance\cite{IK}.

Unfortunately, unlike treating of single-impurity problem of Ref. \cite
{Coleman}, the fields ${\cal Q}$ and ${\cal P}$ can not be completely
removed by gauge transformation. Thus it will be more convenient for us to
work in the gauge where $\alpha _i=0.$

Integrating over $\Psi _{i\alpha }$, expanding the action to second order in 
${\cal P}_{ij},{\cal Q}_{ij}$ at $\alpha _i=0$ we obtain the effective
action in the form 
\begin{eqnarray}
{\cal S}_{eff} &=&-N\text{STr}\ln [G_b^{-1}({\bf q},i\omega _n)G_c^{-1}({\bf %
q},i\omega _n)]-\text{STr}\ln [G_a^{-1}({\bf q},i\omega _n)G_f^{-1}({\bf q}%
,i\omega _n)]  \nonumber \\
&&\ \ \ \ \ \ +\sum_{{\bf q},i\omega _n}\sum_{\delta ,\delta ^{\prime }}%
{\cal R}_{{\bf \delta }}({\bf q},i\omega _n)\left( 
\begin{array}{cc}
\Pi _{{\bf \delta \delta }^{\prime }}^{cb}({\bf q},i\omega _n) & \delta _{%
{\bf \delta \delta }^{\prime }} \\ 
\delta _{{\bf \delta \delta }^{\prime }} & -\Pi _{{\bf \delta \delta }%
^{\prime }}^{af}({\bf q},i\omega _n)
\end{array}
\right) \overline{{\cal R}}_{{\bf \delta }^{\prime }}({\bf q},i\omega _n)
\label{Act} \\
&&\ \ \ \ \ \ -\sum_i(\zeta _iY+\lambda _iQ_0+\mu \delta )  \nonumber
\end{eqnarray}
where ${\cal R}_{{\bf \delta }}=({\cal Q}_{{\bf \delta }},{\cal P}_{{\bf %
\delta }}),$ and the polarization operators are given by 
\begin{eqnarray*}
\Pi _{{\bf \delta \delta }^{\prime }}^{cb}({\bf q},i\omega _n) &=&N\sum_{%
{\bf k},iv_n}G_b({\bf k},i\nu _n)G_c({\bf k+q},i\nu _n+i\omega _n)e^{i{\bf %
k(\delta -\delta }^{\prime })} \\
\ &=&N\sum_{{\bf k}}\frac{n_{{\bf k}}^b+n_{{\bf k+q}}^c}{i\omega
_n-\varepsilon _{{\bf k+q}}^c+\varepsilon _{{\bf k}}^b}e^{i{\bf k(\delta
-\delta }^{\prime })} \\
\Pi _{{\bf \delta \delta }^{\prime }}^{af}({\bf q},i\omega _n) &=&\sum_{{\bf %
k},iv_n}G_f({\bf k},i\nu _n)G_a({\bf k+q},i\nu _n+i\omega _n)e^{i{\bf %
k(\delta -\delta }^{\prime })} \\
\ &=&\sum_{{\bf k}}\frac{n_{{\bf k-q}}^a+n_{{\bf k}}^f}{i\omega
_n-\varepsilon _{{\bf k-q}}^a+\varepsilon _{{\bf k}}^f}e^{i{\bf k(\delta
-\delta }^{\prime })}
\end{eqnarray*}
with $G_{b,c,a,f}({\bf k},i\nu _n)$ are standard Bose (Fermi) Green
functions of corresponding fields with the spectra 
\begin{eqnarray}
\varepsilon _{{\bf k}}^c &=&{\cal A}t_{{\bf k}}+\lambda +\zeta  \nonumber \\
\varepsilon _{{\bf k}}^a &=&{\cal C}t_{{\bf k}}+\lambda +\zeta +\mu 
\nonumber \\
\varepsilon _{{\bf k}}^f &=&{\cal B}t_{{\bf k}}+\lambda -\zeta +\mu 
\nonumber \\
\varepsilon _{{\bf k}}^b &=&{\cal F}t_{{\bf k}}+\lambda -\zeta  \label{ek}
\end{eqnarray}
The mean-field values ${\cal F},{\cal A},{\cal C}$ and ${\cal B}$ have to be
determined self-consistently 
\begin{eqnarray}
{\cal F} &=&-\langle f_i^{\dagger }f_j\rangle ,\;{\cal B}=-\langle
b_{i\sigma }^{\dagger }b_{j\sigma }\rangle  \nonumber \\
{\cal C} &=&\langle c_{i\sigma }^{\dagger }c_{j\sigma }\rangle ,\;{\cal A}%
=\langle a_i^{\dagger }a_j\rangle  \label{FBCA}
\end{eqnarray}
For $\lambda ,\zeta $ and $\mu $ we have the constraint equations 
\begin{eqnarray}
\langle f_i^{\dagger }f_i\rangle +\langle a_i^{\dagger }a_i\rangle &=&\delta
\nonumber \\
\langle b_{i\sigma }^{\dagger }b_{i\sigma }\rangle +\langle c_{i\sigma
}^{\dagger }c_{i\sigma }\rangle &=&1-\delta  \nonumber \\
\langle b_{i\sigma }^{\dagger }b_{i\sigma }\rangle +\langle f_i^{\dagger
}f_i\rangle +\frac 1{Q_0}\langle \theta _i^{\dagger }\theta _i\rangle &=&2S
\label{Constr}
\end{eqnarray}

Further we consider the most interesting case of a ``mixed'' phase where
both the bosons $a,b$ are condensed. Physically, this corresponds to a
non-saturated ferromagnetic state. The then concentrations $\delta _c,\delta
_c^{\prime }$ where the condensates of $b$ and $a$ bosons vanish will
determine the transitions into saturated ferromagnetic and nonmagnetic
states respectively.

First, we consider the standard (noninteracting) mean-field theory which
neglects the fluctuations of the ${\cal Q}$ and ${\cal P}$ fields, i.e. the
second line of (\ref{Act}). We obtain in this case at $T=0$%
\begin{eqnarray}
n_a &=&{\cal A}_0=\delta -\sum_{{\bf k}}n_{{\bf k}}^f  \nonumber \\
n_b &=&-{\cal B}_0=1-\delta -\sum_{{\bf k}}n_{{\bf k}}^c  \label{EqNI}
\end{eqnarray}
where $n_{{\bf k}}^{c,f}=N_F(\varepsilon _{{\bf k}}^{c,f})$, $%
N_F(\varepsilon )$ is the Fermi distribution function, the spectra $%
\varepsilon _{{\bf k}}^f$,$\varepsilon _{{\bf k}}^c$ are determined by (\ref
{ek}) with 
\begin{eqnarray}
\lambda _0 &=&\zeta -{\cal F}_0t_0  \nonumber \\
\lambda _0+\mu _0 &=&-\zeta -{\cal C}_0t_0
\end{eqnarray}
and $n_{a,b}$ are densities of condensates, index $0$ at the parameters
stands for their noninteracting mean-field values. It can be checked
numerically that the equations (\ref{EqNI}) do not have the solutions with
positive $n_a,$ $n_b$ for any $\delta ,\zeta .$

To improve the behavior of solutions of equations (\ref{EqNI}), we consider
the corrections to above noninteracting mean-field theory owing to
fluctuations of ${\cal Q},{\cal P}$ (remember that the fluctuations of $%
\alpha $ were excluded by gauge transformation). Introducing Fourier
components
\begin{eqnarray*}
{\cal Q}_{{\bf q},{\bf \delta }} &=&\sum_i{\cal Q}_{i,i+\delta }e^{i{\bf q}%
R_i} \\
{\cal P}_{{\bf q},{\bf \delta }} &=&\sum_i{\cal P}_{i,i+\delta }e^{i{\bf q}%
R_i}
\end{eqnarray*}
and taking the functional derivatives of action (\ref{Act}), we obtain for
the following expressions 
\begin{eqnarray}
\langle b_{{\bf k}\sigma }^{\dagger }b_{{\bf k}\sigma }\rangle  &=&n_{{\bf k}%
}^b+\sum_{{\bf q},i\omega _n}\left[ \frac 1T\frac{n_{{\bf k}}^b(1+n_{{\bf k}%
}^b)}{i\omega _n-\varepsilon _{{\bf k+q}}^c+\varepsilon _{{\bf k}}^b}+\frac{%
n_{{\bf k}}^b+n_{{\bf k+q}}^c}{(i\omega _n-\varepsilon _{{\bf k+q}%
}^c+\varepsilon _{{\bf k}}^b)^2}\right] G_Q^{\delta \delta ^{\prime }}({\bf q%
},i\omega _n)e^{i{\bf k(\delta -\delta }^{\prime })}  \nonumber \\
&&\ -\frac{\delta {\cal F}t_{{\bf k}}+\delta \lambda }Tn_{{\bf k}}^b(1+n_{%
{\bf k}}^b)  \nonumber \\
\langle c_{{\bf k}\sigma }^{\dagger }c_{{\bf k}\sigma }\rangle  &=&n_{{\bf k}%
}^c+\sum_{{\bf q},i\omega _n}\left[ \frac 1T\frac{n_{{\bf k}}^c(1-n_{{\bf k}%
}^c)}{i\omega _n-\varepsilon _{{\bf k}}^c+\varepsilon _{{\bf k-q}}^b}-\frac{%
n_{{\bf k-q}}^b+n_{{\bf k}}^c}{(i\omega _n-\varepsilon _{{\bf k}%
}^c+\varepsilon _{{\bf k-q}}^b)^2}\right] G_Q^{\delta \delta ^{\prime }}(%
{\bf q},i\omega _n)e^{i({\bf q}-{\bf k)(\delta -\delta }^{\prime })} 
\nonumber \\
&&\ +\frac{\delta {\cal A}t_{{\bf k}}+\delta \lambda }Tn_{{\bf k}}^c(1-n_{%
{\bf k}}^c)  \nonumber \\
\langle a_{{\bf k}}^{\dagger }a_{{\bf k}}\rangle  &=&n_{{\bf k}}^a+\sum_{%
{\bf q},i\omega _n}\left[ \frac 1T\frac{n_{{\bf k}}^a(1+n_{{\bf k}}^a)}{%
i\omega _n-\varepsilon _{{\bf k}}^a+\varepsilon _{{\bf k-q}}^f}+\frac{n_{%
{\bf k}}^a+n_{{\bf k-q}}^f}{(i\omega _n-\varepsilon _{{\bf k}}^a+\varepsilon
_{{\bf k-q}}^f)^2}\right] G_P^{\delta \delta ^{\prime }}({\bf q},i\omega
_n)e^{-i{\bf k(\delta -\delta }^{\prime })}  \nonumber \\
&&\ -\frac{\delta {\cal C}t_{{\bf k}}+\delta \lambda +\delta \mu }Tn_{{\bf k}%
}^a(1+n_{{\bf k}}^a)  \nonumber \\
\langle f_{{\bf k}}^{\dagger }f_{{\bf k}}\rangle  &=&n_{{\bf k}}^f+\sum_{%
{\bf q},i\omega _n}\left[ \frac 1T\frac{n_{{\bf k}}^f(1-n_{{\bf k}}^f)}{%
i\omega _n-\varepsilon _{{\bf k+q}}^a+\varepsilon _{{\bf k}}^f}-\frac{n_{%
{\bf k}}^f+n_{{\bf k+q}}^a}{(i\omega _n-\varepsilon _{{\bf k+q}%
}^c+\varepsilon _{{\bf k}}^b)^2}\right] G_P^{\delta \delta ^{\prime }}({\bf q%
},i\omega _n)e^{i({\bf k+q)(\delta -\delta }^{\prime })}  \label{Av} \\
&&\ +\frac{\delta {\cal B}t_{{\bf k}}+\delta \lambda +\delta \mu }Tn_{{\bf k}%
}^f(1-n_{{\bf k}}^f)  \nonumber
\end{eqnarray}
and 
\begin{equation}
\langle \theta _i^{\dagger }\theta _i\rangle =-\sum_{{\bf q},i\omega
_n}\left[ \Pi _{{\bf \delta \delta }^{\prime }}^{cb}({\bf q},i\omega _n)-\Pi
_{{\bf \delta \delta }^{\prime }}^{af}({\bf q},i\omega _n)\right] 
\label{Avt}
\end{equation}
where $\delta {\cal F},\delta {\cal A},\delta {\cal C}$,$\delta {\cal B}%
,\delta \lambda $ and $\delta \mu $ denote the corrections to corresponding
quantities of noninteracting mean-field theory owing to interaction with the
fields ${\cal Q}$ and ${\cal P};$ the Green functions $G_{Q,P}^{\delta
\delta ^{\prime }}$ are defined by
\begin{eqnarray*}
G_Q^{\delta \delta ^{\prime }}({\bf q},\tau ) &=&\langle T[{\cal Q}_{{\bf q},%
{\bf \delta }}(\tau )\overline{{\cal Q}}_{{\bf q},{\bf \delta }^{\prime
}}(0)]\rangle  \\
G_P^{\delta \delta ^{\prime }}({\bf q},\tau ) &=&\langle T[{\cal P}_{{\bf q},%
{\bf \delta }}(\tau )\overline{{\cal P}}_{{\bf q},{\bf \delta }^{\prime
}}(0)]\rangle 
\end{eqnarray*}
and can be found by inverting corresponding matrix in (\ref{Act}). To keep
the spectrum of the bosons $a,b$ gapless we choose 
\begin{eqnarray}
\delta \lambda  &=&\sum_{{\bf q},i\omega _n}\frac 1{i\omega _n-\varepsilon _{%
{\bf q}}^c}G_Q^{\delta \delta ^{\prime }}({\bf q},i\omega _n)-\delta {\cal F}%
t_0  \nonumber \\
\delta \lambda +\delta \mu  &=&\sum_{{\bf q},i\omega _n}\frac 1{i\omega
_n-\varepsilon _{-{\bf q}}^f}G_P^{\delta \delta ^{\prime }}({\bf q},i\omega
_n)-\delta {\cal C}t_0  \label{dlm}
\end{eqnarray}
The expressions (\ref{Av}) and (\ref{Avt}) together with the self-consistent
equations (\ref{FBCA}), (\ref{Constr}) give the parameters of supersymmetric
mean-field theory.

\section{Conclusion}

In this paper we have obtained the supersymmetric action of $t-J$ model,
which interpolates between slave-fermion and slave-boson ones. We have also
developed a mean-field approximation for this action, which takes into
account the fluctuations owing to the interaction with gauge fields.
Physically, these fluctuations correspond to fluctuations of the total boson
(fermion) number in the system, since only total number of bosons and
fermions is conserved in the supersymmetric representation\cite{Coleman}. In
other words, these interactions describe virtual boson-fermion transitions
within the supersymmetric fermion-boson representation. The account of these
virtual transitions can give in principle a possibility to investigate the
crossover from the saturated ferromagnetic state into nonmagnetic one and.
After a generalization to finite $U,$ the crossovers from antiferromagnetic
state to the states which are described well by slave-boson representation,
e.g. pseudo-gap and superconducting one, could be described. The application
of the results of the present paper to solving these problems is the aim of
future work.

\section{Appendix. Gauge transformation of mean-field Hamiltonian parameters
and Green functions}

In this Appendix we consider the transformation laws for different
quantities under the gauge transformation (\ref{G}). We treat $\eta _i(\tau
) $ as a sort of fluctuations (see, e.g. \cite{BLP}) with zero average and
the pair correlation function 
\begin{equation}
f({\bf q},\omega )=\langle \overline{\eta }({\bf q},\omega )\eta ({\bf q}%
,\omega )\rangle
\end{equation}
Calculating the contributions up to quadratic terms in $\eta $, we obtain
following transformation laws of the Hamiltonian parameters: 
\begin{eqnarray}
\lambda &\rightarrow &\lambda -\sum_{{\bf q},i\omega _n}(2\zeta -i\omega
_n)f({\bf q},i\omega _n)  \nonumber \\
{\cal F} &\rightarrow &{\cal F}-\sum_{{\bf q},i\omega _n}({\cal F}-{\cal A}%
t_{{\bf q}}/t_0)f({\bf q},i\omega _n)  \nonumber \\
{\cal A} &\rightarrow &{\cal A}+\sum_{{\bf q},i\omega _n}({\cal A}-{\cal F}%
t_{{\bf q}}/t_0)f({\bf q},i\omega _n)  \nonumber \\
{\cal B} &\rightarrow &{\cal B}-\sum_{{\bf q},i\omega _n}({\cal B}-{\cal C}%
t_{{\bf q}}/t_0)f({\bf q},i\omega _n)  \nonumber \\
{\cal C} &\rightarrow &{\cal C}+\sum_{{\bf q},i\omega _n}({\cal C}-{\cal B}%
t_{{\bf q}}/t_0)f({\bf q},i\omega _n)  \label{tr1}
\end{eqnarray}
The gauge fields are transformed as 
\begin{eqnarray}
\alpha _i &\rightarrow &\alpha _i+(\partial _\tau +2\zeta _i)\eta _i 
\nonumber \\
{\cal Q}_{ij} &\rightarrow &{\cal Q}_{ij}+{\cal A}\eta _i-{\cal F}\eta _j 
\nonumber \\
{\cal P}_{ij} &\rightarrow &{\cal P}_{ij}+{\cal C}\eta _i-{\cal B}\eta _j
\label{trg}
\end{eqnarray}
Using (\ref{trg}) one can obtain the transformation laws for the gauge
fields Green functions, e.g., 
\begin{eqnarray}
G_\alpha ({\bf q},i\omega _n) &\rightarrow &G_\alpha ({\bf q},i\omega
_n)+(2\zeta -i\omega _n)^2f({\bf q},i\omega _n)  \label{tr2} \\
G_Q^{\delta \delta ^{\prime }}({\bf q},i\omega _n) &\rightarrow &G_Q({\bf q}%
,i\omega _n)+({\cal A}-{\cal F}e^{i{\bf q\delta }})({\cal A}-{\cal F}e^{i%
{\bf q\delta }^{\prime }})f({\bf q},i\omega _n)  \nonumber \\
G_P^{\delta \delta ^{\prime }}({\bf q},i\omega _n) &\rightarrow &G_Q({\bf q}%
,i\omega _n)+({\cal C}-{\cal B}e^{i{\bf q\delta }})({\cal C}-{\cal B}e^{i%
{\bf q\delta }^{\prime }})f({\bf q},i\omega _n)  \nonumber
\end{eqnarray}
The transformation laws for other quantities can be obtained by combining
the results (\ref{tr1}), (\ref{tr2}).

\end{document}